\documentclass[10pt]{article}

\usepackage{graphicx}
\usepackage{color} 

\usepackage{amsmath}
\usepackage{amsfonts}
\usepackage{amssymb}
\usepackage{siunitx}
  
\begin{document}

\title{CRC 1114 -- Report \\
Membrane Deformation by N-BAR Proteins: \\
Extraction of membrane geometry and protein diffusion characteristics from MD simulations}

\author{J.\ H.\ Peters, C. Gr\"aser, R.\ Klein}

\maketitle

\section{Membrane Shaping Proteins (N-BAR domain)}
\label{sec:n-bar}

A good overview of mechanisms shaping biological membranes can be found in \cite{jarsch_membrane_2016}. In a previous study (\cite{arkhipov_four-scale_2008}), membrane bending by a BAR domain was investigating using four models of different resolutions.

\subsection{Simulation Setup}

Membrane protein structures are notoriously difficult to determine and hence only a small number of structures are available, most of which are of lower quality. The N-BAR system used here was based on two crystallographic structures, a lower quality dimer structure (PDB id 4I1Q \cite{sanchez-barrena_bin2_2012}) and a higher quality monomer structure (PDB id: 4AVM \cite{allerston_crystal_nodate}). While the monomer structure is of higher quality than the dimer one, it still is missing a number of atoms, which were using PDBFixer (https://github.com/pandegroup/pdbfixer accessed 30.10.2017) but also a N-Terminal alpha-helix which is assumed to be essential in membrane binding (\cite{gallop_mechanism_2006,arkhipov_four-scale_2008}) and which had to be modelled using PyMOL (https://pymol.org). Two copies of the corrected monomer structure were then fitted on the two chains of the dimer structure to produce a fixed dimer structure to be used in simulation.

To reduce interaction between periodic images, the dimer was placed in contact with a big (600\AA x 600\AA) membrane patch. As in previous work \cite{arkhipov_four-scale_2008}, the membrane was composed of $70\%$ DOPC and $30\%$ DOPS which carries a net charge of $-1e$ per molecule, as BAR-domain membrane interactions are likely electrostatic in nature (\cite{arkhipov_four-scale_2008}). To reduce computational cost, the MARTINI coarse-grained forcefield \cite{monticelli_martini_2008,de_jong_improved_2012} was used. The system was set up using CHARMM-GUI (http://www.charmm-gui.org) \cite{jo_charmm-gui:_2008,qi_charmm-gui_2015,hsu_charmm-gui_2017}. And simulated using GROMACS 5.1.2 \cite{abraham_gromacs:_2015}.

After equilibration, a total of 300 ns were simulated using suggested simulation parameters from CHARMM-GUI: A time step of 20 fs was used and neighbour searching was performed every 10 steps. The Shift algorithm was used for electrostatic interactions with a cut-off of 1.2 nm. A single cut-off of 1.4 was used for Van der Waals interactions. The V-rescale temperature coupling algorithm was used to keep the system at 303.15K and semi-isotropic pressure coupling was done with the Berendsen algorithm.

\subsection{Simulation results}

\subsubsection{2D diffusion of N-BAR dimer}

One possible long-term goal of the collaboration with project C07 of CRC 1114 was the design of a dynamic model of proteins diffusing (under consideration of membrane-induced interaction potentials) on a membrane. To this end, the 2D diffusion coefficient of a N-BAR Dimer on a membrane is of interest.

\begin{figure}[h]
  \includegraphics[width=\textwidth]{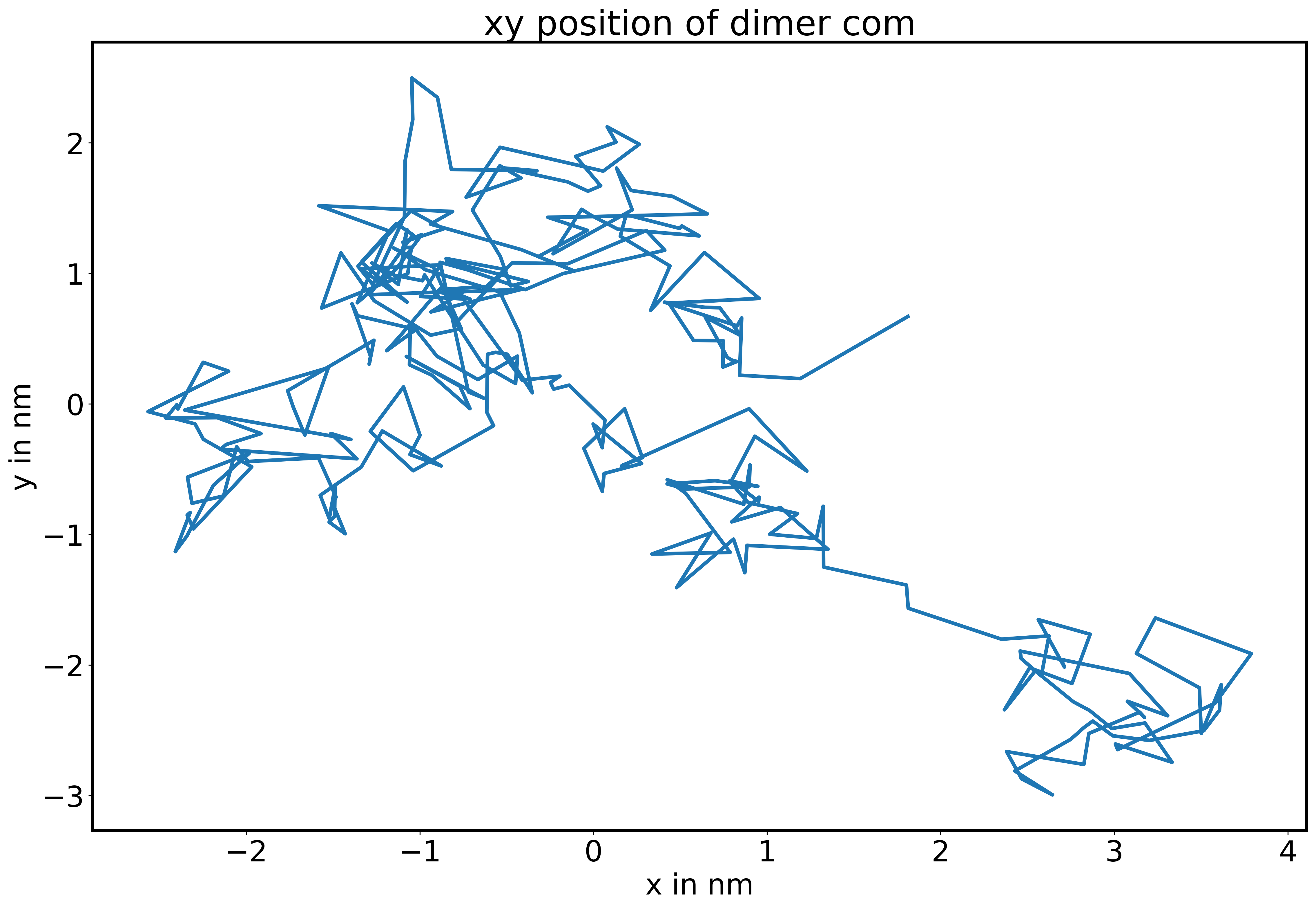} 
  \caption{\label{fig:DimerXYTrajectory}Trajectory of the center of mass of the Dimer projected to the XY plane.}
\end{figure}

\begin{figure}[h]
  \includegraphics[width=\textwidth]{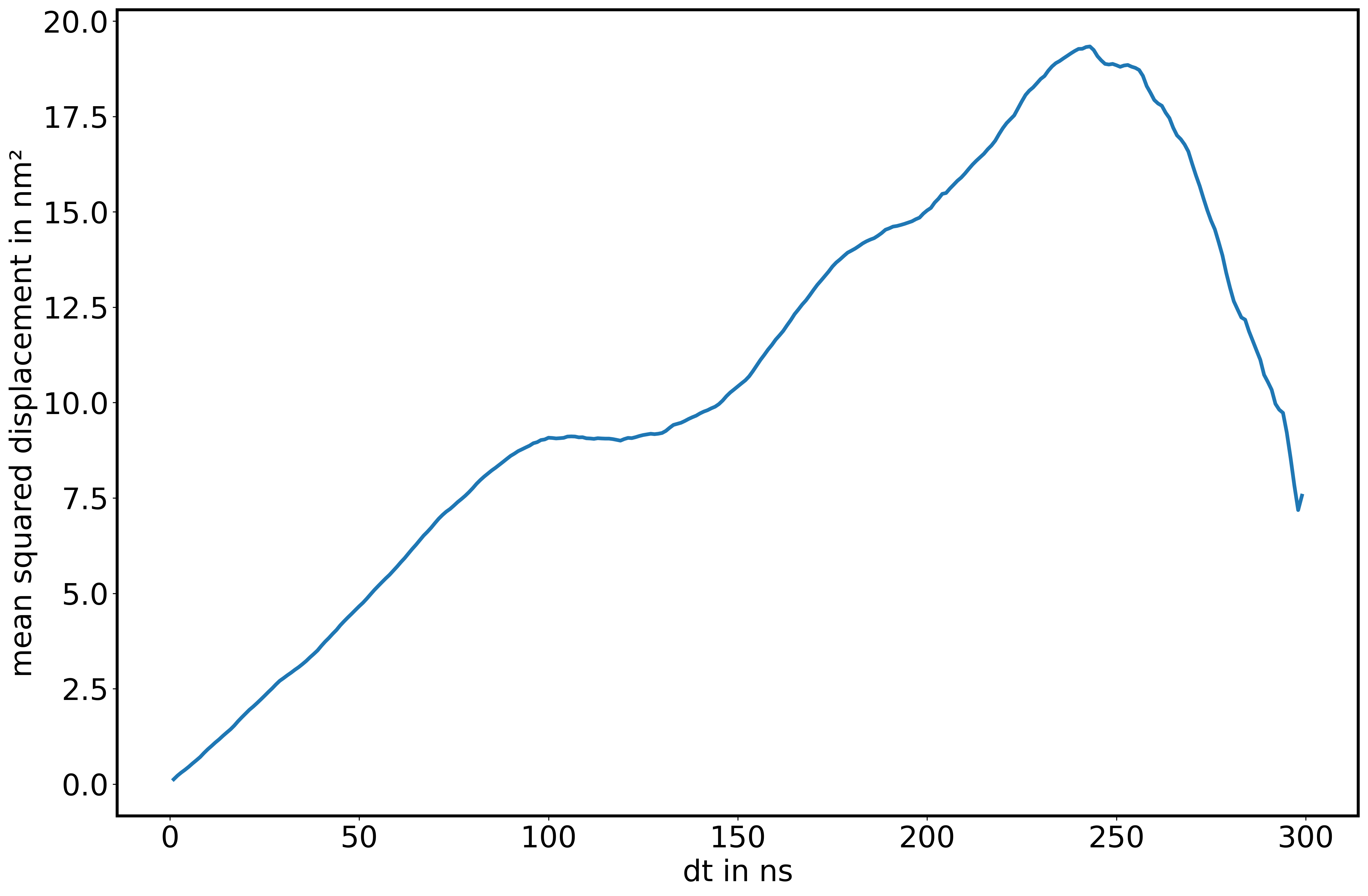} 
  \caption{\label{fig:DimerXYMSD}Mean squared displacement (in the XY-plane) of the dimer center of mass for different time differences.  For $dt \leq 100 ns$ statistics are good enough to be described as normal diffusion.}
\end{figure}

The trajectory of the centre of mass of the N-BAR dimer from 1 ns snapshots (figure \ref{fig:DimerXYTrajectory}) corresponds to a random walk on a 2D plane defined by the membrane. Calculating from this the mean squared displacement over time (figure \ref{fig:DimerXYMSD}),the diffusion constant can be approximated using 

$$\langle x^2 \rangle = dDt$$

\begin{figure}[h]
  \includegraphics[width=\textwidth]{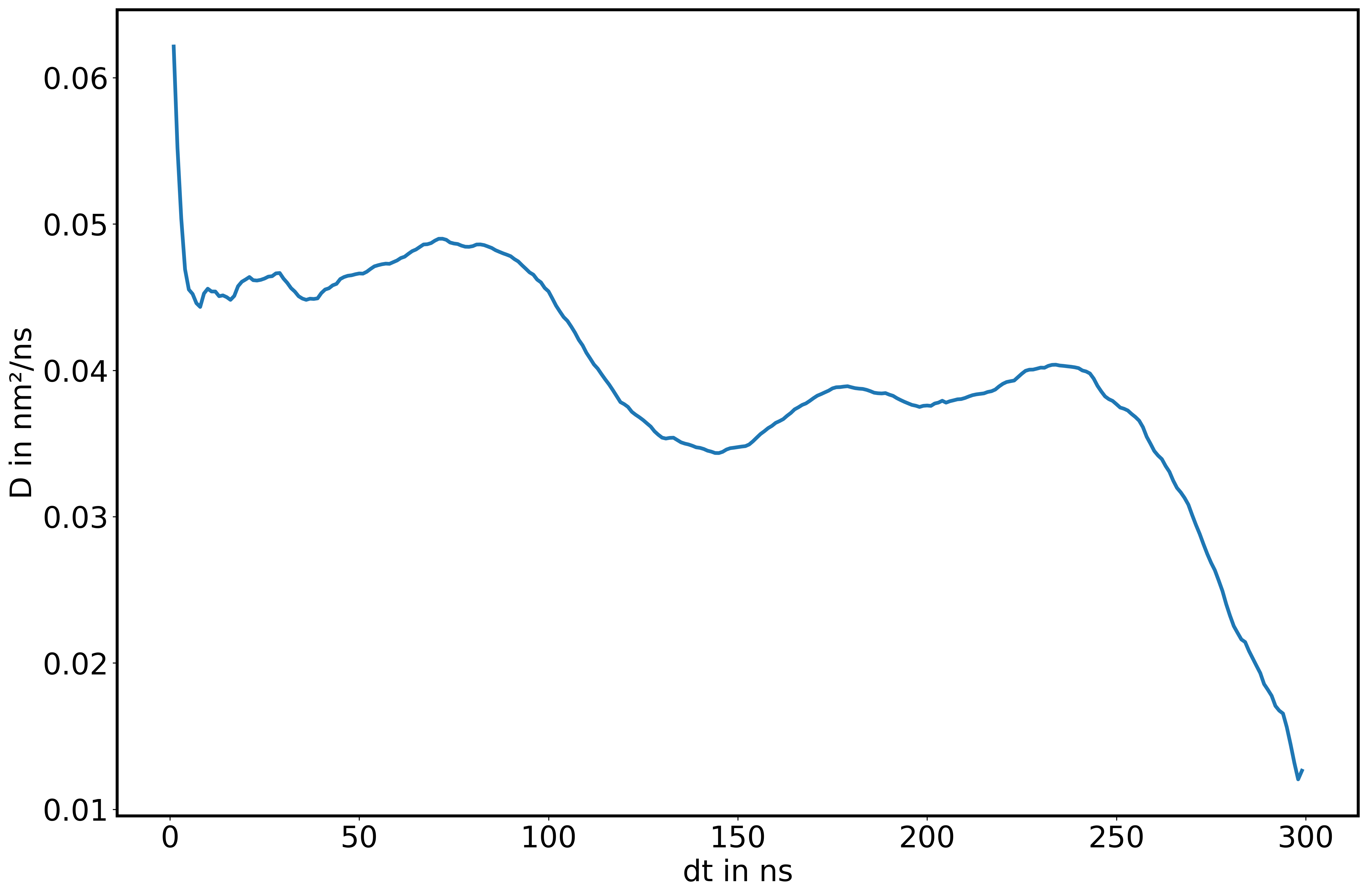} 
  \caption{\label{fig:DimerXYD}2D diffusion coefficient calculated for different maximal time differences.}
\end{figure}

Based on $dt < 100$ns (the regime for which $\langle x^2 \rangle$ increases linear over time as would be expected in normal diffusion, figure \ref{fig:DimerXYMSD}), the diffusion coefficient is $D = (0.047 \pm 0.002) nm^2/ns$. Using all data (including the low statistics $dt > 250$ ns), $D = (0.039 \pm 0.008) nm^2/ns$ 

\subsubsection{Induced curvature}

\begin{figure}[h]
  \includegraphics[width=0.475\textwidth]{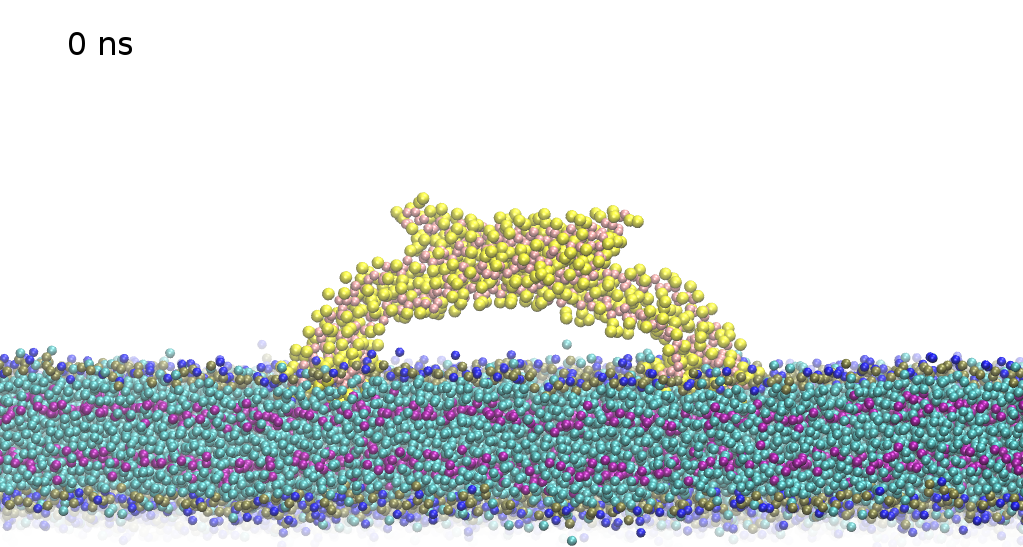} 
  \includegraphics[width=0.475\textwidth]{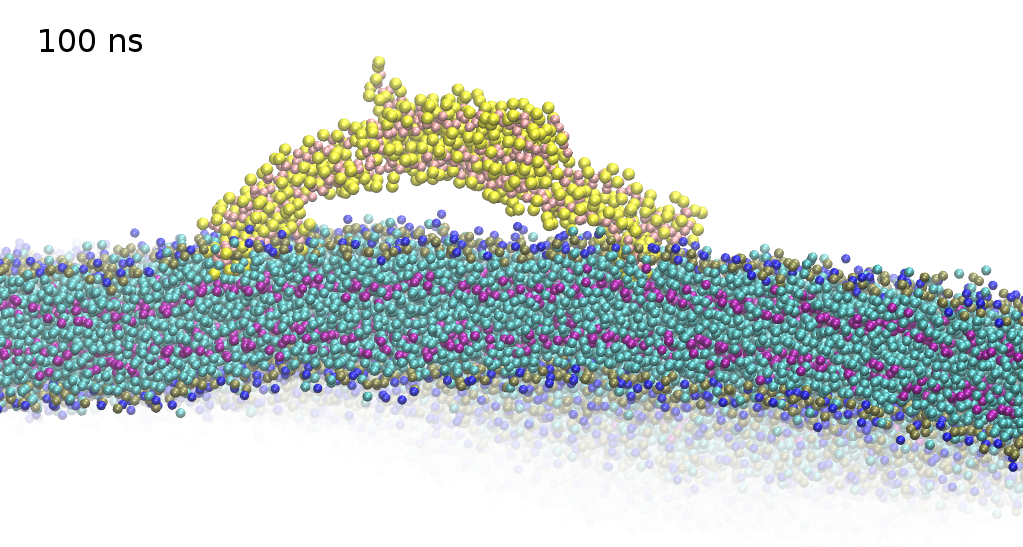} 
  \includegraphics[width=0.475\textwidth]{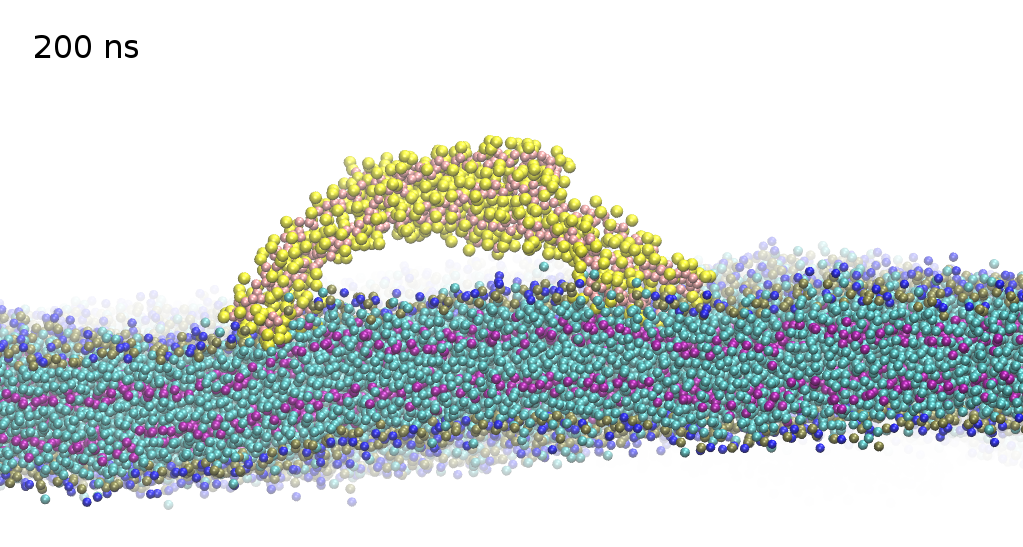} 
  \includegraphics[width=0.475\textwidth]{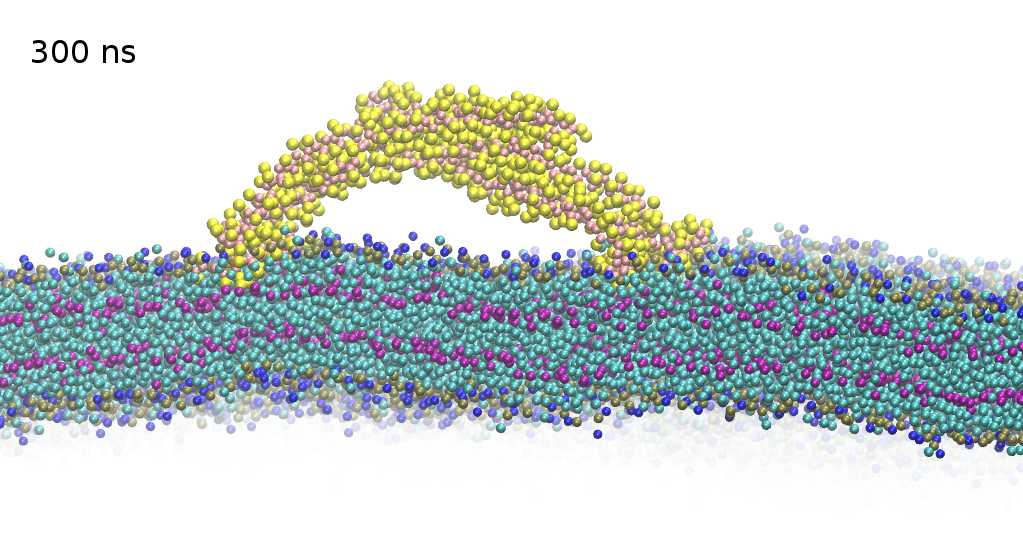} 
  \caption{\label{fig:N-BAR_structure}Membrane deformation by N-BAR Dimer.}
\end{figure}

During the whole simulation time, the N-BAR dimer remains attached to the membrane (figure \ref{fig:N-BAR_structure}).

\begin{figure}[h]
  \includegraphics[width=\textwidth]{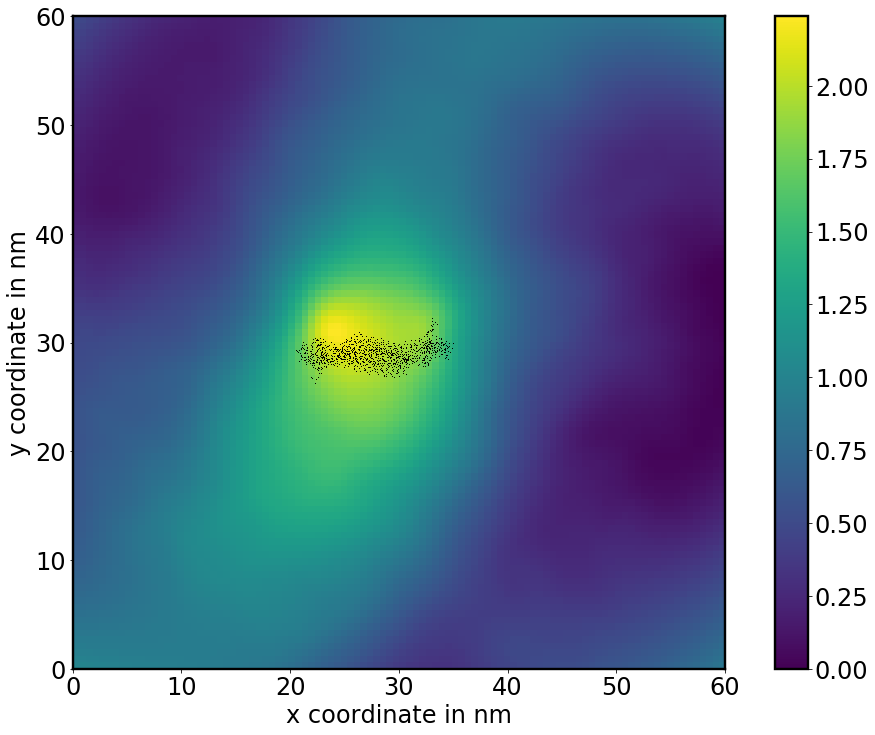} 
  \caption{\label{fig:N-BAR_Membrane_Deformation}Average deformation (for trajectory from 100 ns to 300 ns) of the membrane due to interaction with an N-BAR dimer (indicated in black).}
\end{figure}

\begin{figure}[h]
  \includegraphics[width=\textwidth]{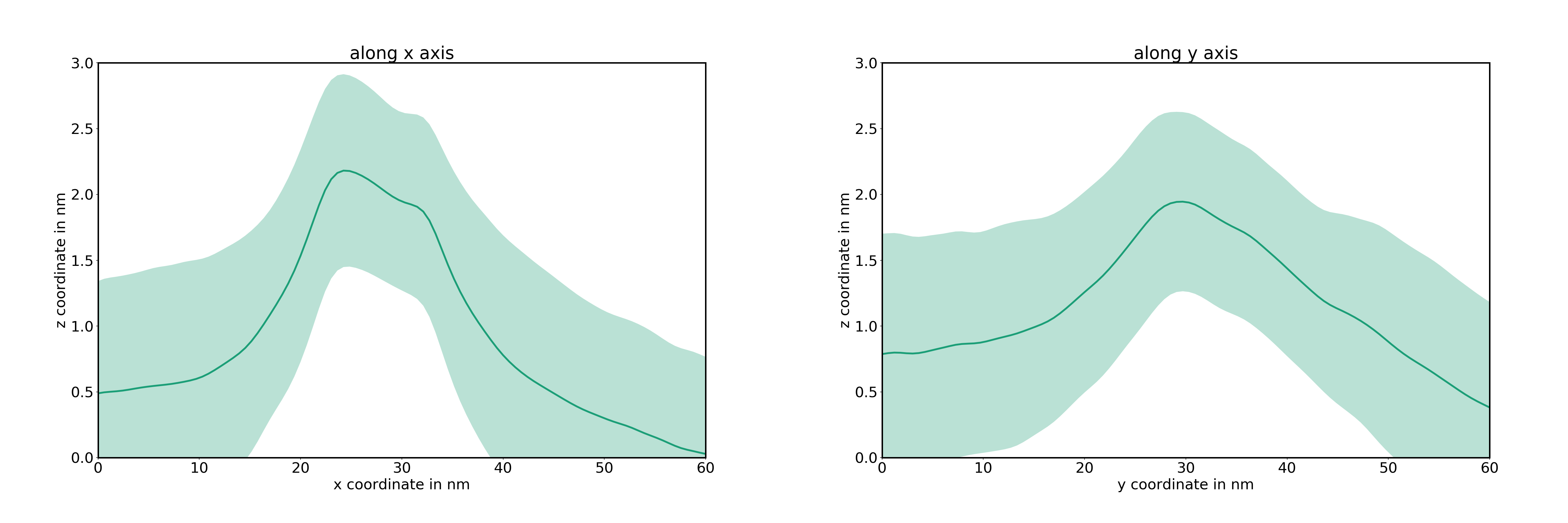} 
  \caption{\label{fig:N-BAR_Membrane_height}Average height profile of the membrane in x-direction and y-direction through the N-BAR dimer position. The shaded areas describe the standard deviation of membrane fluctuations over time.}
\end{figure}

To quantify the effect of the N-BAR dimer, the curvature of the membrane at the location of the protein has to be calculated. Since the membrane fluctuates over time (figure \ref{fig:N-BAR_Membrane_height}), an average over a long time (200 snapshots representing 200 ns simulation time) results in a clear average membrane deformation (figure \ref{fig:N-BAR_Membrane_Deformation}). To calculate the average, translational movement in the xy plane and rotational movement around the z-axis of the N-BAR dimer had to be removed.

\clearpage
\section{Geometric Membrane Insertions}
\label{sec:cones}

There are only a small number of fully resolved structure of molecules known to shape biological lipid membranes, and even fewer of transmembrane proteins known to have this effect. For a more systematic investigation of the effect cone-shaped membrane insertions, we are using geometrically defined pseudo-molecules.

As protein domains that are inserted into biological membranes usually have corresponding electrostatic properties (i.e. they have charged or hydrophilic residues where the protein is in contact with charged or hydrophilic lipid head groups and hydrophobic where the protein is in contact with hydrophobic lipid tails), such insertions ideally should have similar properties.

\begin{figure}[h]
  \includegraphics[width=\textwidth]{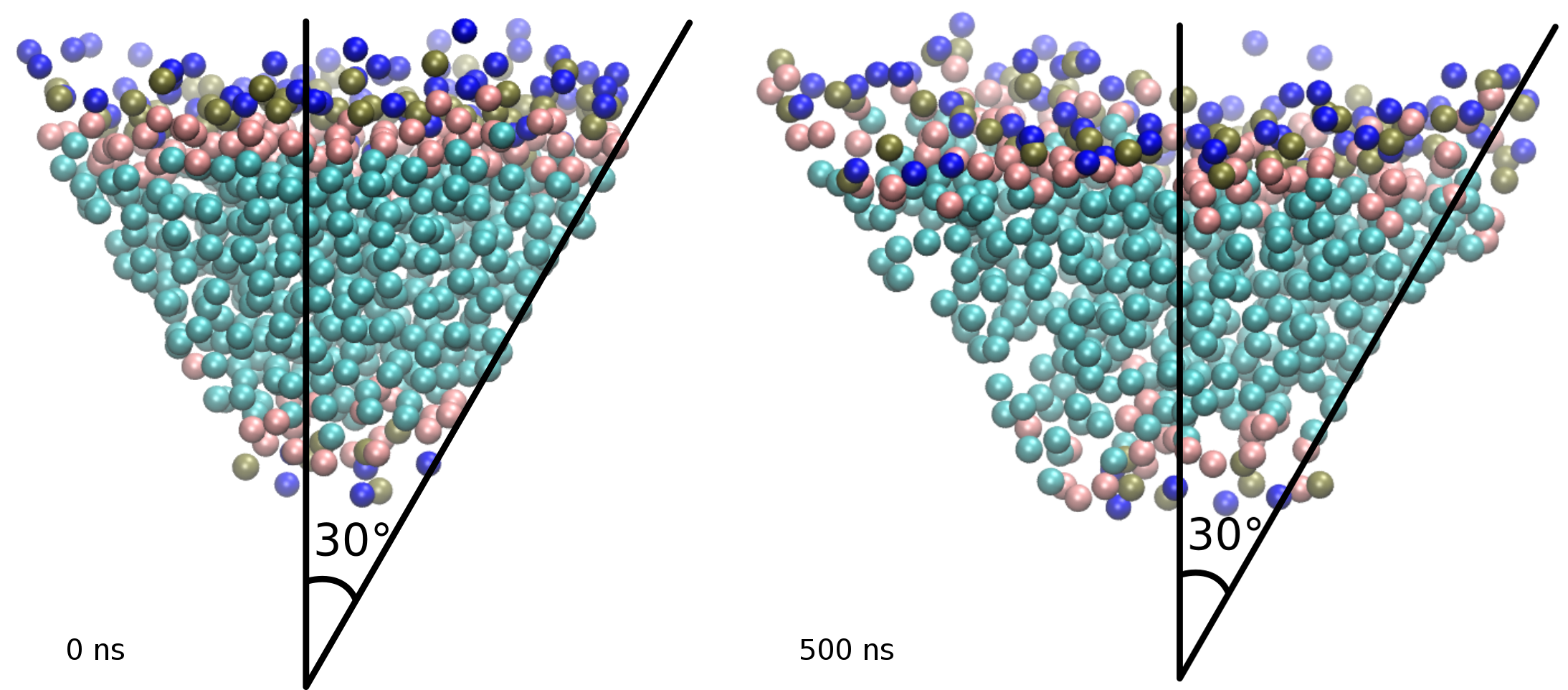} 
  \caption{\label{fig:cone}Cone shaped membrane insertion with an opening angle of $\ang{30}$. When position restraints are used, the cone does retain its initial shape (left). By using nearest neighbour bonds and bond angles, the cone can be kept at approximately the same angle even after longer simulation time (right). In this case, the cone can move through the simulation box.}
\end{figure}

One simple way to construct such domains is to select atoms from an existing lipid bilayer structure selected to be located along the desired geometry (figure \ref{fig:cone}).

In a simulation, these artificial trans-membrane domains are inserted into a lipid bilayer domain consisting of DPPC lipids. To have a mobile but stable cone structure, bonds and bond angles between the atoms have been introduced to form cone-shaped "pseudo molecules" (figure \ref{fig:cone}).

As there is no explicit influence of water in these interactions, we have chosen DryMARTINI \cite{arnarez_dry_2015}, the implicit solvent variant of the MARTINI coarse grained forcefield for these simulations.

A total of 500 ns were simulated with a time step of 20 fs. Neighbour searching was performed every 10 steps. The Shift algorithm was used for electrostatic interactions with a cut-off of 1.2 nm and a single cut-off of 1.4 was used for Van der Waals interactions. Semi-isotropic pressure coupling was done with the Berendsen algorithm. As is common for simulations using the DryMARTINI forcefield, the leap-frog stochastic dynamics integrator implemented in GROMACS was used with a reference temperature of 323K (DPPC transitions into the gel phase below 314K) and an inverse friction constant of 4.0 ps.

\begin{figure}[h]  
  \includegraphics[width=\textwidth]{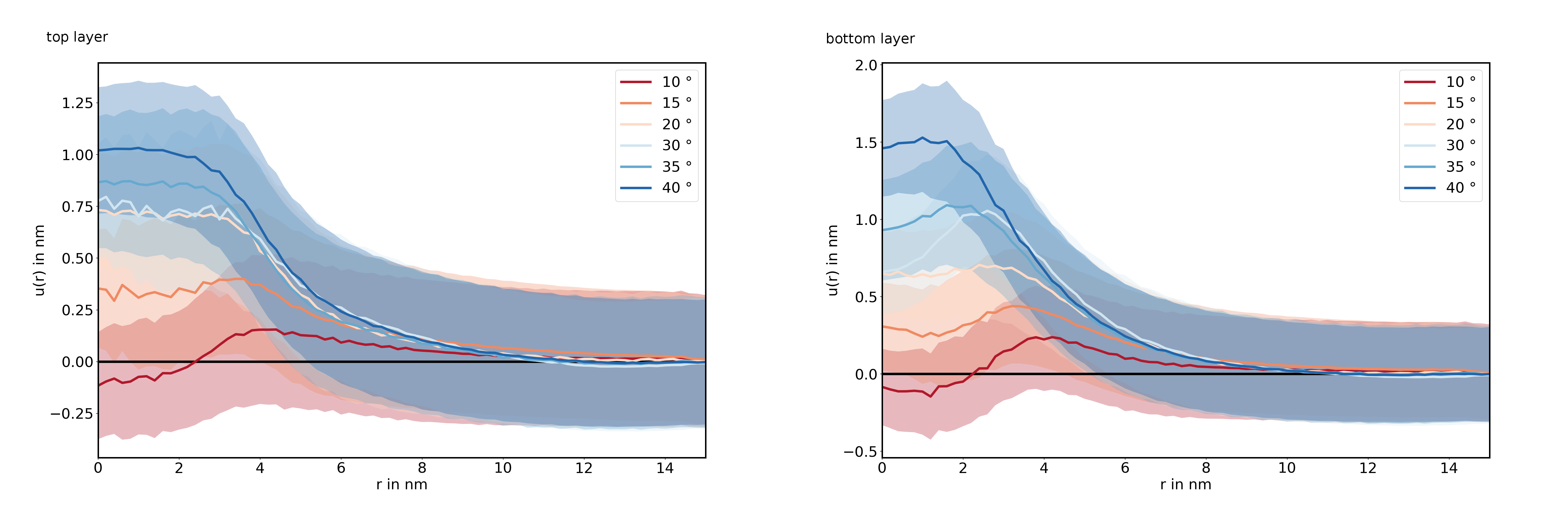}
  \caption{\label{fig:CONE_Membrane_Deformation}Average deformation of the membrane due to interaction with a cone-shaped  insertion. the shaded areas describe the standard deviation of the membrane fluctuations.}
\end{figure}

Membrane insertions are known to have a weaker effect on membrane curvature than memberane-attached proteins the BAR domain (\cite{jarsch_membrane_2016}). Again, systematic membrane deformation can best be observed by averaging over longer trajectories. Also, since the membrane insertions are rotationally symmetric, the height profile was averaged radially (figure \ref{fig:CONE_Membrane_Deformation}).

\section{Mapping from Molecular Dynamics to Continuum Model}
\label{sec:Rbf}

Eliott Et. Al. \cite{elliott_variational_2016} have developed a variational approach to model the interaction between membranes and membrane-shaping particles based on the Canham-Helfrich continuum model \cite{canham_minimum_1970,helfrich_elastic_1973,evans_bending_1974}, which requires the bending rigidity to describe the response of the membrane to interaction.

Both for membrane stiffness calculations (section \ref{sec:membranestiffness}) and for comparison with continuum models, it is helpful to map the particle representation of the membrane to a continuous model of the form $h(x,y) = z$. Radial basis functions are a mesh-free model of the form

$$h(x,y) = \sum_n^N c_n f(\sqrt{(x-x_n)^2+(y-y_n)^2}) $$

By using the coordinates of $N$ atoms $(x_n,y_n,z_n)$ as anchor points of the rbf, the function can be constructed in a way to satisfy

$$h(x_n,y_n) = z_n$$

for all points by selecting the correct $c_n$ through a simple set of linear equations.

To account for statistical fluctuations in atom positions, we divide the membrane into a number of layers (using the same atom from each lipid molecule), construct an rbf for each of them (after centring them in z-direction) and then average over these models.

\clearpage
\section{Bending Stiffness Calculations}
\label{sec:membranestiffness}

An approach to calculate the membrane bending stiffness from equilibrium simulations (in contrast to non-equilibrium simulations like vesicle closing simulaitons) has been developed by Khelashvili Et. Al. (\cite{khelashvili_calculating_2013,johner_implementation_2016}). While the original implementation (\cite{johner_implementation_2016} requires a solvent density to calculate the membrane surface (and from it the surface normal which is required for the calculation of the bending stiffness), our Rbf-based membrane fitting approach can also be applied to implicit solvent systems where no water density is available.

Following the method described in \cite{johner_implementation_2016}, we have calculated bending rigidity of an unperturbed DOPC/DOPS membrane using the explicit solvent MARTINI coarse grained forcefield (corresponding to the membrane used in the N-BAR domain simulations, section \ref{sec:n-bar}) and a DPPC membrane using the implicit solvent DryMARTINI forcefield (corresponding to the membrane used in the geometric insertion simulations, section \ref{sec:cones} ). The resulting bilayer bending moduli are listed in table  \ref{tab:bendingRigidity}.

\begin{table}[h]
  \begin{center}
    \begin{tabular}{| l  r |}
      \hline
      membrane system & bending rigidity \\ \hline
      DOPC/DOPS & $(16.8 \pm 0.8) k_BT$\\
      DPPC      & $(20.2 \pm 0.8) k_BT$\\ 
      \hline
    \end{tabular}
    \caption{\label{tab:bendingRigidity} Bending rigidities calculated for unperturbed membrane systems}
  \end{center}
\end{table}

\clearpage

\section{Conclusions}

\section*{Acknowledgements}
This research has been funded by Deutsche Forschungsgemeinschaft (DFG) through grant CRC 1114, project C01. The authors acknowledge the North-German Supercomputing Alliance (HLRN) for providing HPC resources that have contributed to the research results reported in this paper. We would like to thank Prof. Roland Netz for his suggestions leading to the simulation of cone-shaped membrane inclusions.

\clearpage

\bibliographystyle{ieeetr} 
\bibliography{membrane_shaping}

\end{document}